\documentstyle[a4,axodraw]{article}

\newcommand\order{{\cal O}}

\renewcommand\Im{\mbox{Im}}

\newcommand\GEV{\mbox{GeV}}

\newcommand\PB{\mbox{pb}}
\newcommand{\gsim}{\stackrel{\lower.7ex\hbox{$>$}}
        {\lower.7ex\hbox{$\sim$}}}
\def\setepsfscale#1{\def\epsfsize##1##2{#1##1}}
\setepsfscale{.5}
\def\axoscale{.5 }
\begin{document}
\title{UNSTABLE PARTICLES AND GAUGE INVARIANCE}
\author{Geert Jan van Oldenborgh\\
{\em Paul Scherrer Institute, CH-5232 Villigen PSI, Switzerland}\\}
\maketitle
\setlength{\baselineskip}{2.6ex}
\begin{abstract}
\footnotesize\noindent
When computing the properties of reactions involving unstable charged
particles care has to be taken to use a gauge invariant amplitude.  In this
talk we present methods to (automatically) obtain such an amplitude, both at
the tree level and in one-loop calculations, using only a minimal number
of diagrams.  The numerical difference with gauge variant methods commonly
used will be discussed for the case of tree level W pair production.
\end{abstract}
\maketitle
\begin{flushright}
PSI-PR-93-19
\end{flushright}

%###] titlepage:
%###[ Introduction:

\section{Introduction}

Many of the cross sections one would like to evaluate with the methods
described elsewhere in these proceedings involve the production and decay of
unstable particles.  The $W$ and $Z$ bosons both have a sizable width $\Gamma$
($\Gamma/m\approx1/40$), and the same may be true for the top quark and Higgs
boson (depending on their masses).  For these particles the well-known narrow
width approximation will not be sufficient for many purposes.  Including
off-shell effects, however, involves a resummation of some of the higher order
graphs, which means that care has to be taken to keep the matrix element gauge
invariant.  This last point is not only of academic interest as gauge
breaking terms are often much larger numerically than the gauge invariant
result.  Therefore even suppression factors $\Gamma/m$ may not be enough to
avoid inaccuracies.

The Dyson resummation of higher order graphs in the propagator, which leads to
a finite width, is clearly only needed when the unstable particle can
kinematically be on its mass shell.  Indeed, for an unstable particle with
space-like momentum $p^2<0$ the imaginary part of the self energy is zero,
hence no finite width should be used.  One can therefore not use a complex
mass in the Lagrangian, even disregarding the problems with the standard model
mass generation mechanism which would follow (for instance a complex value for
$\sin^2\theta_W$).  It is a purely kinematical problem which particles should
be given a finite width and which not.

The amplitude can thus be divided into a set of {\em resonant\/} diagrams,
which contain a divergence within the allowed phase space as an unstable
particle is taken on-shell, and {\em non-resonant\/} graphs without such
divergences.  In principle the contribution of the non-resonant diagrams is
suppressed with respect to the resonant ones by $\Gamma/m\propto\alpha$.  We
first discuss ways to only use the resonant tree graphs to get the dominant
behaviour, next increasing the accuracy of the result by including the
non-resonant tree graphs and the resonant one-loop graphs.  A more detailed
discussion will be given in a forthcoming publication
\cite{Andre&Geert&Daniel}.  The tree level results are based on Ref.\
\cite{OGamma}.

%###] Introduction:
%###[ Resonant Tree Graphs:

\section{Resonant Tree Graphs}

In this section some schemes will be given to deal with the resonant tree
graphs.  For simplicity these are first discussed for a single unstable
particle with momentum $p$ and mass $m$; the extension to more particles will
be given later later.  We assume that the diagram generator has already sorted
out which diagrams ones are resonant and which are not.  Without resummation
the full (gauge invariant) amplitude for this process therefore has the form
\begin{equation}
\label{eq:naive}
    {\cal M} = \frac{R(p^2)}{p^2 - m^2} + N(p^2)
\ .
\end{equation}
The essential variable here is the virtuality $p^2$ of the unstable particle.
The other kinematical variables are assumed to be CMS angles, which are
independent of $p^2$.  Unfortunately, this amplitude gives an infinite cross
section when the integration region includes the pole at $p^2 = m^2$.

The simplest approximation one can make is to completely factorize the process
in production and decay: the narrow width approximation.  This fails to take
into account terms of $\order(\Gamma)$; the corresponding dimensionless
parameter will usually be $\Gamma/m$ (about 1/40 for the $W$), but can be much
larger near the threshold for production of the unstable particles.

The most important of these $\order(\Gamma)$ corrections are included by
simply using the on-shell expression for the matrix element, but treating the
kinematics and unstable particle propagator off-shell.  This off-shell
propagator is derived by resumming the one-loop corrections to the propagator,
giving rise to a simple Breit-Wigner propagator $1/(p^2-m^2 + im\Gamma)$ with
$\Gamma$ defined by the relation $m\Gamma = \Im\Pi(m^2)$ in terms of the self
energy $\Pi$ evaluated at the real mass.  We refer to this procedure as the
``resonant'' scheme.  This procedure obviously respects gauge invariance and
will serve as the base line to compare the others with.

The next step is usually to also evaluate the matrix element off-shell, we
refer to this procedure as the ``off-shell'' method.  This also gives an
answer below threshold.  In the case of a charged resonance however, this
procedure violates $U(1)_{em}$ gauge invariance, even if the width is given by
the physical (on-shell) quantity, which is gauge invariant. The reason is that
the original resonant graphs  are not the only graphs which lead to the
particular final state:  non-resonant graphs have to be included.

%###] Resonant Tree Graphs:
%###[ Non-resonant Tree Graphs:

\section{Non-resonant Tree Graphs}

One reason to include the non-resonant graphs is to conserve gauge invariance.
Another is that they can be sizable due to some other near-divergence, such as
large collinear logarithms.  The ``naive'' way to include these non-resonant
graphs is to just add them to the resonant terms with a complex mass.
However, as the amplitude was gauge invariant for $\Gamma=0$ it follows that
it must be gauge variant for $\Gamma > 0$ (if the resonant graphs are not
separately gauge invariant).

An obviously gauge invariant procedure to include the non-resonant graphs was
introduced by Zeppenfeld {\em et al}\ \cite{Zeppenfeld&Co}.  The price to save
gauge invariance in this ``overall'' scheme is an incorrect treatment of the
non-resonant contribution close to mass shell.   However, it can be argued
that the difference is of higher order in this region of phase space.  We thus
have a prescription which is correct to leading order both on resonance and
away from it, but these contributions are formally of different order in
$\alpha$.

Another gauge invariant way to include the off-shell effects is to
systematically separate orders in $\Gamma$\cite{Stuart1}.  As the residue at
the pole $p^2=m^2$ is gauge invariant one can add the finite width in the
first term without breaking gauge invariance.  This corresponds to adding and
resumming only the gauge invariant part of the propagator corrections.  In
this scheme the cross section is given as the sum of the resonant cross
section plus $\order(\Gamma)$ corrections, which are both gauge invariant.  It
is thus a natural starting point for higher order corrections.

However, this ``polescheme'' also has some undesirable properties.  The first
(resonant) term has a discontinuity when the threshold for the production of
the unstable particle is crossed.  Approaching from below, one even encounters
the original (non-resummed) singularity in the propagator.  The accuracy of
this scheme is thus doubtful around threshold.

The difference between the last three schemes is of order $\Gamma^2$ for total
cross sections and singly differential distributions, but of order $\Gamma$
for doubly and higher differential distributions.   This follows from the
occurrence of a Levi-Civita tensor in the $\order(\Gamma)$ terms.  A summary
of all schemes is given in table \ref{table}.

\begin{table}[tb]
\centerline{%\small
\begin{tabular}{|p{1.5cm}|l|l|l|l|p{2.3cm}|}
\hline
name & kine-  & matrix element & gauge   & treatment & threshold \\
     & matics &                & invari- & non-resonant & behaviour \\
     &        &                & ant?     & terms      & \\
\hline
{\raggedright narrow width} & on-shell & $\displaystyle R(m^2)$ & yes & none &
{\raggedright undefined be-\\low threshold}\\
resonant & off-shell & $\displaystyle
\frac{R(m^2)}{p^2-m^2+im\Gamma}$ & yes & none & {\raggedright zero below\\
threshold}\\
off-shell & off-shell & $\displaystyle
\frac{R(p^2)}{p^2-m^2+im\Gamma}$ & no  & none & {\raggedright wrong below\\
threshold}\\
naive    & off-shell &
$\displaystyle\frac{R(p^2)}{p^2-m^2+im\Gamma} + N(p^2)$ & no  & ok & ok
\\[5mm]
overall  & off-shell &
$\begin{array}[t]{l}\displaystyle\frac{R(p^2)}{p^2-m^2+im\Gamma}\\\displaystyle
        \mbox{} + \frac{p^2-m^2}{p^2-m^2+im\Gamma}N(p^2)\!\!\!\end{array}$ &
yes & wrong & ok
\\[8mm]
pole\-scheme & off-shell &
$\begin{array}[t]{l}\displaystyle\frac{R(m^2)}{p^2-m^2+im\Gamma}\\
        \mbox{} + \displaystyle\frac{R(p^2) - R(m^2)}
            {p^2 - m^2}\\
        \mbox{} + N(p^2)\end{array}$ &
yes & ok & {\raggedright wrong around\\ threshold}\\
\hline
\end{tabular}
}
\caption{The six different schemes to treat unstable particles at tree level.}
\label{table}
\end{table}

The extensions to the case where there are more unstable particles are largely
straightforward.  There now appears a hierarchy of $n$-fold resonant diagrams.
The first three methods just use the maximally resonant ones.  In the
``polescheme'' one only needs these and the ones where only one particle is
non-resonant; the ``overall'' scheme uses all diagrams.  The threshold
behaviour of the `polescheme'' now is worse than before.  However, in
this area the worth of any fixed-order calculation is doubtful because of
possible bound state effects.

\begin{figure}[tb]
\begin{center}
\begin{picture}(444,444)(75,40)
\put(0,0){\strut\epsffile{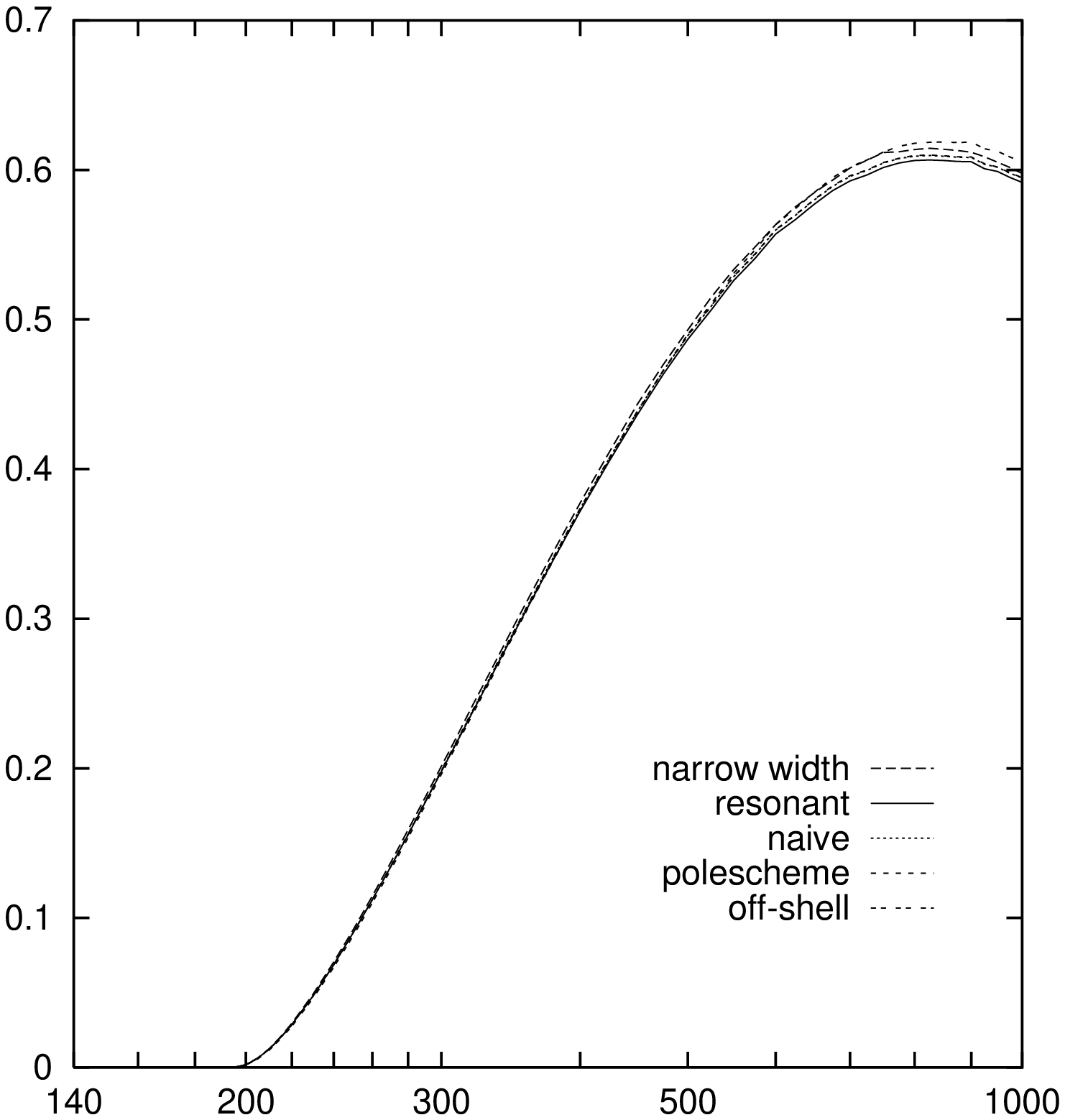}}
\put(100.8,488.9){\makebox(0,0)[bl]{$\sigma[\PB]$}}
\put(480.8, 19.1){\makebox(0,0)[tr]{$\sqrt{s_{ee}}[\GEV]$}}
\end{picture}\nobreak
\begin{picture}(444,444)(75,40)
\put(0,0){\strut\epsffile{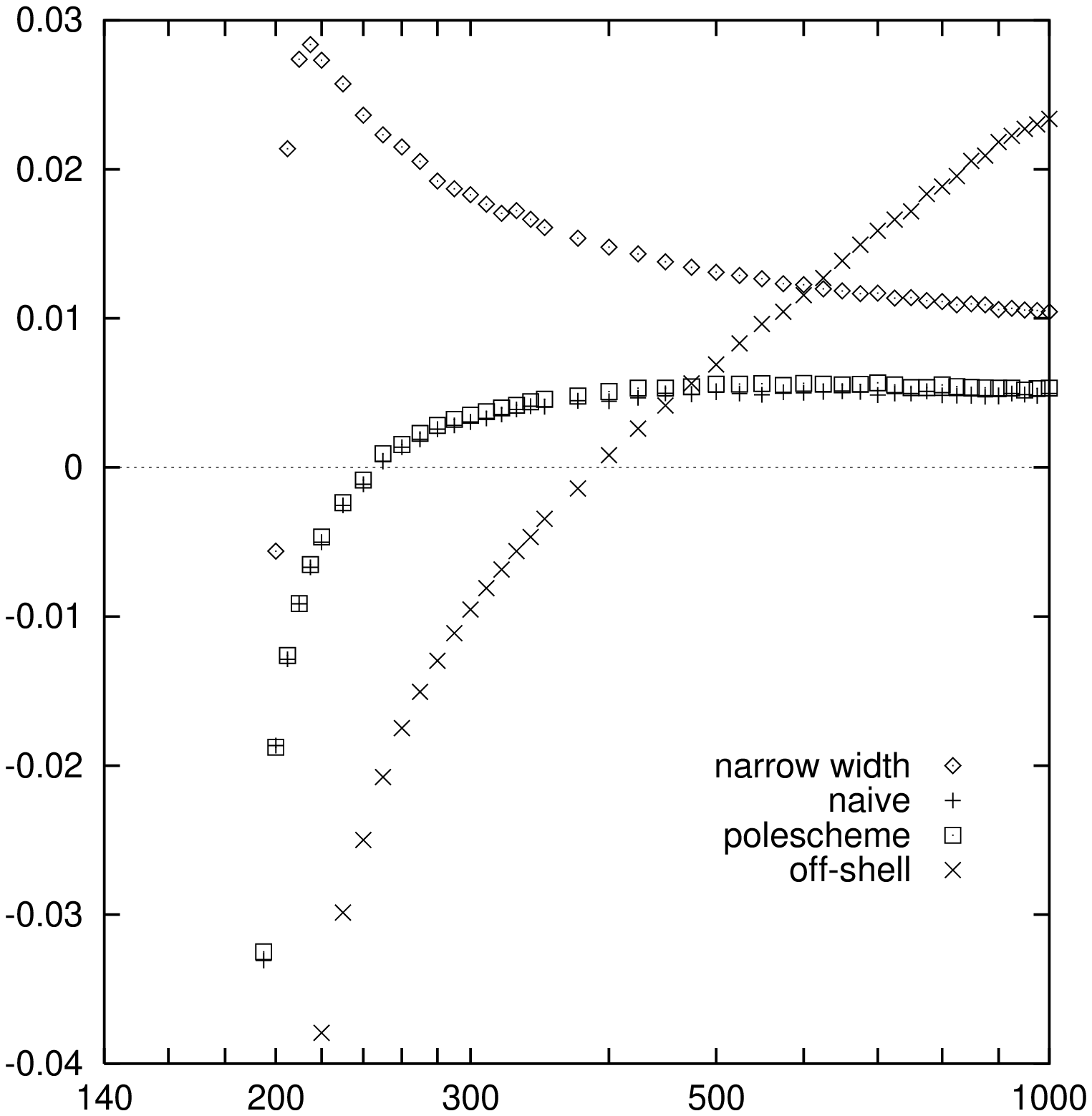}}
\put(100.8,488.9){\makebox(0,0)[bl]{$\frac{\displaystyle
\sigma-\sigma_{\rm res}}{\displaystyle\sigma_{\rm res}}$}}
\put(480.8, 19.1){\makebox(0,0)[tr]{$\sqrt{s_{ee}}[\GEV]$}}
\end{picture}
\end{center}
\caption[]{Cross sections for $\gamma\gamma\to \ell^- \bar{\nu}_\ell W^+$ in
the different schemes defined in the text
with $p_\perp(\ell) > 0.02\sqrt{s_{ee}}$}
\label{fig:polyWW}
\end{figure}

The numerical influence on the total cross section is illustrated in Fig.\
\ref{fig:polyWW}, where the total cross section for
$\gamma\gamma\to \ell^- \bar{\nu}_\ell W^+$ is plotted,
with the $W^+$ considered stable
and a photon beam derived from backscattered electrons \cite{Ginsburg}.  The
physically relevant off-shell effects will be twice as large.  To avoid the
collinear divergence in the final state a minimum transverse
momentum $p_\perp(\ell) > 0.02 \sqrt{s}$ is demanded;
this should also give an indication of detector acceptance.
It can be seen that the narrow width approximation misses some important
effects.  The graph for the offshell scheme is misleading, in that we used the
formula $\sum_i \epsilon^\mu_i \epsilon^\nu_i = -g^{\mu\nu}$ which is not
valid when the photon couples to a non-conserved current.  It does show that
one has to be careful when breaking gauge invariance.  One should also note
that the differences between the various schemes are expected to be much
larger in doubly differential distributions.

%###] Non-resonant Tree Graphs:
%###[ Resonant One-loop Corrections:

\section{Resonant One-loop Corrections}

When computing the $\order(\alpha)$ corrections one is normally only
interested in the resonant part.  We will thus have to identify which
diagrams contribute to this, and then to find a gauge invariant way to isolate
just the resonant part.  A generalization of the ``polescheme'' is proposed for
this last step.  In the case of a neutral resonance (like the $Z$ boson) the
isolation step is not necessary, as the resonant diagrams already form a gauge
invariant subset; the only bone of contention left in this case amounts to the
proper definition of the mass of the unstable particle.

There are three classes of resonant diagrams.  The first one contains the
corrections to the resonant propagator.  A gauge invariant part of these are
resummed to give the finite width; the rest (vanishing at $p^2=m^2$) has to be
treated perturbatively.  The second group, the factorizable diagrams,
comprises the corrections to either production or decay of an unstable
particle; these diagrams obviously retain the original resonant propagators.
Added to this, charged resonances will give rise to logarithms $\log(p^2-m^2)$
as the virtuality and finite width of the unstable particle regulate the
infrared divergence.  The third class of non-factorizable diagrams is given by
the set of infrared divergent diagrams with the photon (or gluon) spanning one
or more resonant propagators.   Simple power counting arguments show that no
other non-factorizable diagrams can give a resonant contribution.  One would
suspect that even the contributions from these diagrams may cancel against the
corresponding soft Bremsstrahlung diagrams.  As the hard Bremsstrahlung is
easily seen to be non-resonant this would imply the intuitively satisfying
result that initial-final and final-final state interaction vanish in the
narrow width approximation.

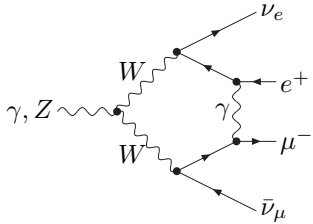
\begin{figure}[htb]
\unitlength .75bp
\def\axoscale{.75 }
\begin{center}
\begin{picture}(120,70)(0,30)
\Photon(0,60)(30,60){2}{3}
\Photon(30,60)(60,90){2}{5}
\Photon(30,60)(60,30){2}{5}
\ArrowLine(110,75)(90,75)
\ArrowLine(90,75)(60,90)
\ArrowLine(60,90)(100,110)
\ArrowLine(100, 10)(60,30)
\ArrowLine(60,30)(90,45)
\ArrowLine(90,45)(110,45)
\Photon(90,45)(90,75){2}{3}
\Vertex(30,60){2}
\Vertex(60,90){2}
\Vertex(60,30){2}
\Vertex(90,45){2}
\Vertex(90,75){2}
\put(-2,60){\makebox(0,0)[r]{$\gamma,Z$}}
\put(45,76){\makebox(0,0)[br]{$W$}}
\put(45,43){\makebox(0,0)[tr]{$W$}}
\put(112, 75){\makebox(0,0)[l]{$e^+$}}
\put(102,110){\makebox(0,0)[l]{$\nu_e$}}
\put(112, 45){\makebox(0,0)[l]{$\mu^-$}}
\put(102, 10){\makebox(0,0)[l]{$\bar{\nu}_\mu$}}
\put(87, 60){\makebox(0,0)[r]{$\gamma$}}
\end{picture}
\end{center}
\caption{A double resonant five point function occurring in $W$ pair
production}
\label{fig:E0}
\end{figure}

However, this is not the case.  As an example consider the doubly resonant
five point function shown in Fig.\ \ref{fig:E0}.   Using the Schouten identity
one can decompose it in five four point functions as
\begin{equation}
E_0 = \frac{-1}{2\Delta_{5s}}\left( \Delta_{ps}^\gamma D_0(\gamma) +
\Delta_{ps}^\mu D_0(\mu) + \Delta_{ps}^{W^-} D_0(W^-) + \Delta_{ps}^{W^+}
D_0(W^+) + \Delta_{ps}^{e} D_0(e) \right)
\end{equation}
where $D_0(p)$ denotes the scalar four point function with the propagator
corresponding to particle $p$ missing and the $\Delta$'s denote determinants
of external and internal momenta \cite{gjthesis}.  The overall factor
$-1/2\Delta_{5s}$ is quadratic in the virtualities $p_\pm^2 - m^2$ of the $W$
bosons and thus gives rise to the double resonance.  The $D_0(\gamma)$ and
$\Delta_{ps}^\gamma$ are both finite, whereas for the other combinations the
determinant is linear in the virtuality and the four point function linearly
divergent.

The corresponding scalar Bremsstrahlung integral can be decomposed
similarly, except that the equivalent of the $D_0(\gamma)$ never
appears.  The other four terms exactly match the corresponding terms in the
virtual diagram, leaving this one doubly resonant term.  There is no symmetry
which would cause it to disappear, even when integrated over phase space, and
there are no other graphs which will give rise to the same four point
function.  It therefore is a double resonant term of order $\alpha$ which is
not suppressed by additional factors $\Gamma/m$.  The failing cancellations
can be traced back to the fact that, in the limit $\Gamma\ll\lambda$
($\lambda$ is an infrared regulator, normally the photon mass) the linear and
quadratic infrared divergences do {\em not\/} cancel.  Only a numerical study
can decide whether the non-factorizable diagrams give rise to
measurable effects.

The diagrams we have selected as resonant diagrams in general do not form a
gauge invariant subset.  One can select a gauge invariant part by taking the
residue at the pole, as was done in the previous sections.  Unfortunately,
this residue is ill-defined due to the occurrence of the divergent logarithms.
(These also do not cancel between the virtual and Bremsstrahlung graphs.)
These logarithms only occur in the scalar functions however.  As the
coefficients of the scalar functions are gauge invariant, one can evaluate
these any way one likes.

The following recipe therefore serves as the extension of the ``polescheme''
for one-loop (and soft Bremsstrahlung) diagrams:
\begin{list}{$\bullet$}{\itemsep 0pt plus 0pt minus 0pt}
\item The reduction of the diagram to scalar functions should be done on-shell
(if the unstable particle is charged).
\item The scalar functions can be evaluated in any convenient non-divergent
way; on-shell or off-shell.  This also allows one to take into account the
finite width in non-divergent but non-analytical logarithms like the Coulomb
singularity and threshold effects.
\end{list}
This schemes have been implemented in the FF library \cite{FFguide}.

%###] Resonant One-loop Corrections:
%###[ Conclusions:

\section{Conclusion}

In this talk we have presented a scheme to systematically include off-shell
effects in amplitudes including unstable particles, explicitly keeping gauge
invariance.  Unfortunately it breaks down near the threshold for the
production of the unstable particles.  In this region, however, the occurrence
of bound state effects makes any fixed-order perturbation series suspect.  The
numerical difference of this procedure with the (gauge variant) ones commonly
used are small in the tree level total cross section, at least in the case of
$W$ pair production and the unitary gauge.  Larger differences are expected
for doubly differential cross sections.

%###] Conclusions:
%###[ postamble:

\section{Acknowledgements}

The results presented here are based on work done together with Daniel
Wyler and Andre Aeppli in the context of the one-loop corrections to $W$ pair
production.

%\bibliographystyle{nucphys}
%\bibliography{fenomeen}

\begin{thebibliography}{1}

\bibitem{Andre&Geert&Daniel}
A.~Aeppli, G.~J. van Oldenborgh and D.~Wyler.
\newblock  (to appear).

\bibitem{OGamma}
A.~Aeppli, F.~Cuypers and G.~J. van Oldenborgh.
\newblock Phys.\ Lett. {\bf B314}  (1993) 413.

\bibitem{Zeppenfeld&Co}
D.~Zeppenfeld, J.~A.~M. Vermaseren and U.~Baur.
\newblock Nucl.\ Phys. {\bf B375}  (1992) 3.

\bibitem{Stuart1}
R.~G. Stuart.
\newblock Phys.~Lett. {\bf B262}  (1991) 113,
\newblock Phys.~Lett. {\bf B272}  (1991) 353.

\bibitem{Ginsburg}
I.~F. Ginzburg, Kotkin~G. L., V.~G. Serbo and V.~I. Telnov.
\newblock Nucl.\ Instr. Meth. {\bf 205}  (1983) 47.

\bibitem{gjthesis}
G.~J. van Oldenborgh.
\newblock One-loop Calculations with Massive Particles.
\newblock Ph.D. thesis, Universiteit van Amsterdam, 1991.

\bibitem{FFguide}
G.~J. van Oldenborgh.
\newblock Comp.~Phys.~Comm. {\bf 66}  (1991) 1.

\end{thebibliography}

\end{document}